\documentclass[useAMS,usenatbib,usegraphicx]{mn2e}

\title[Measuring transverse velocities in gravitationally lensed systems]{Measuring transverse velocities in gravitationally lensed extragalactic systems using an annual parallax effect}

\author[Tuntsov, Walker \& Lewis]{
A. V. Tuntsov\thanks{E-mail: tyomich@physics.usyd.edu.au}, 
M. A. Walker\thanks{E-mail: maw@physics.usyd.edu.au} \&
G. F. Lewis\thanks{E-mail: gfl@physics.usyd.edu.au},\\
School of Physics A29, University of Sydney, NSW 2006, Australia}

\begin{document}

\date{Accepted 2004 April 6. Received 2004 March 25; in original form 2004 February 16}

\pagerange{\pageref{firstpage}--\pageref{lastpage}} \pubyear{2004}

\maketitle
\begin{abstract}
A parallax method to determine transverse velocity in
a gravitationally lensed system is described. Using the
annual motion of the Earth around the Sun allows us to
probe the local structure of the magnification map that,
under certain assumptions, can be used to infer the effective
transverse velocity. The method is applied to OGLE data for
QSO2237+0305 and the velocity value is estimated to be
about $15\pm 10\,\rmn{km}\,\rmn{s}^{-1}$ if attributed to the
lensing galaxy or about $420\pm 300\,\rmn{km}\,\rmn{s}^{-1}$
if attributed to the quasar. We find this estimate unreasonably
small and conclude that we have not measured a parallax
effect. We give a short list of properties that a system should
possess to allow a successful implementation of this method.
\end{abstract}


\section{Introduction}

\label{firstpage}

Among the various parameters needed to specify a microlensing model of a lensed system one of the most important is the overall transverse velocity. This is required in order to determine physical parameters from the observed temporal properties. Besides gravitational lensing, measuring the transverse velocities of galaxies is of considerable interest for studies of large-scale structure and kinematics \citep{dekelreview, bernardeaureview}.

There have been a number of successful attempts to use the annual motion of the Earth around the Sun for measuring the transverse velocities of microlenses in our own Galaxy. The microlensing optical depth remains very low within the Local Group and, wherever double lenses are not involved, can be well  described by a Schwarzschild lens model. \citet{gould92} developed a convenient formalism to describe the `annual parallax' effect in this case, that allows one to extract information on physical parameters of the lensing configuration, breaking some of the degeneracies inherent to the classical Paczy\'nski light curve \citep{paczynski}. The first microlensing event which showed a strong annual parallax signal was detected by the MACHO collaboration in the first-year data of the group's Galactic bulge program \citep{alcock95} and six more were found among the longest events over the next seven years; two of these are currently among the best stellar mass black hole candidates \citep{bennett02}. Events with an annual parallax signal were also found by other collaborations -- EROS/PLANET \citep{an}, MOA \citep{bond} and OGLE \citep{ogle99, ogle01, ogledb}.  Perhaps the most spectacular results related to this effect are the detection of the first multi-peaked microlensing event by OGLE collaboration \citep{oglemultipeak}, and direct observations of a lens based on parameters determined via this method \citep{alcock01}.

Much effort has also been put into determination of the degeneracies present in the annual parallax effect description itself. It has been found that constant acceleration of the lens or the source can sometimes mimic the parallax signal \citep{smp}, and there also exists a discrete degeneracy between jerk and parallax parameters that becomes a continuous one when the accuracy of observations decreases \citep{gould04, park}. These degeneracies arise, basically, from the symmetry of the Schwarzschild lens that mixes different kinematic effects into a single geometric parameter -- the distance between the source and the lens expressed in the units of the Einstein-Chwolson radius (e.g., \citealt{schneiderbook}). Some of these degeneracies are broken when the lens is a double object \citep{an}, and as lens becomes more complex the above mentioned radius loses its unique status in the description of the lensing event. The single Schwarzschild lens approximation is, however, applicable to most of the microlensing events in the Galaxy due to the low optical depth to microlensing in our neighbourhood.

In contrast, microlensing in the images of strongly lensed quasars necessarily takes place in regions of high optical depth where the structure of the magnification map is complex and unknown, effectively rendering it impossible to model the light curve in detail. In addition, the intrinsic variability that quasars are expected to possess may be another contaminant of any microlensing signal present in the images' light curves.

For the particular multiply imaged quasar QSO2237+0305 \citep{huchra} in which the microlensing phenomenon was observed for the first time \citep{irwin}, a number of attempts have been made to determine the transverse velocity of the system. The small time delay differences between the four images -- less than a day \citep{schneiderondelays} -- means that the observed fluctuations, uncorrelated
between the images, are dominated by microlensing, with negligible intrinsic variations.

\citet{wyithe} introduced a method to determine the transverse velocity based on the statistics of time derivatives of the microlensing-induced flux variations. Although their method seems to be the only one that makes it feasible to take into account proper motion of the microlenses, a number of parameters must be specified in order to make the velocity estimate, including the microlensing mass spectrum, which, in effect, translates temporal quantities into spatial ones. Under the assumptions made in their paper they find the transverse velocity to be less than $500\,\rmn{km}\,\rmn{s}^{-1}$ (at 95\% confidence level) favouring, depending on the model, values in the range $60\,\rmn{km}\,\rmn{s}^{-1}$ to $400\,\rmn{km}\,\rmn{s}^{-1}$ -- a tighter constraint compared to previously assumed value of $\sim 600\,\rmn{km}\,\rmn{s}^{-1}$ \citep{wittmao}.

Another approach to estimating the effective velocity is to compare the spatial extent of `quiescent' regions of the magnification map models with the temporal extent of the periods of steady rise and fall in the actual light curves of the quasar images -- mostly, image D \citep{quiescent}. However, all these methods are seriously dependent on the assumed microlensing parameters and deal with statistical properties of the light curves, which are difficult to establish with the currently available data.

In this paper we implement another, rather simple approach, underlain by a few natural assumptions. The essence of the method is the following: the light curve of an image of the lensed quasar is composed of the values of magnification on the observer's path through the magnification map (observer's plane). When the region of the map considered is small enough and is far from caustic curves, it is natural to expand the magnification as a function of the observer's position into a Taylor series and restrict ourselves to its linear terms. Where this approximation holds, the well-known motion of the Earth around the Sun can be used to obtain the local values of magnitude and direction of the gradient on this magnification map. Combined with the measured time derivative, they can be used to estimate the velocity of the Sun with respect to magnification map. We apply this method to the data available for QSO2237+0305.

In this analysis it is hardly possible to incorporate proper motions of the lenses and we will also neglect the intrinsic variability of the quasar; therefore it is unreasonable to expect a full accounting of all the observed features. However, as a (nearly) model-independent estimate of the velocity, this approach seems to be an interesting application of the microlensing phenomenon which could be extended to different microlensed systems if their properties are favourable.

In the next section we describe the method we use for probing the structure of the magnification map and obtaining the velocity estimate in greater detail, in section 3 the method is applied to the observational data for QSO2237+0305 obtained by the OGLE collaboration \citep{oglehuchra} and the results are discussed: the value for transverse velocity obtained seems to us to be too low and we conclude that the method has failed in this case. Discussion of desirable properties of lensed systems, which may permit successful application of this method, concludes this study in section 4.

\section{Basic approach}

Flux variations in different images of lensed quasars may be caused by either intrinsic mechanisms, such as accretion instabilities, disc precession etc. (e.g., \citealt{pringle, kpwl}), or by mechanisms related to the propagation of light from the quasar to the observer, which, in the case of lensed quasars are dominated by microlensing in the macrolensing galaxy. The latter itself could be split into two kinds of variations -- projected motion of the observer with respect to a fixed point in the image plane and `proper' variations in the magnification map itself caused by the proper motions of the microlenses.

In the case of an Earth-bound observer the projected motion in turn consists of the constant motion of the Solar system and annual motion of the Earth around the Sun. This `annual parallax' effect in connection with microlensing of quasars was first considered by~\citet{grieger} and was shown to be detectable under certain conditions. However, the authors noted that with the data then available it was unlikely that the effect could be detected and therefore they did not elaborate on its potential applications; for determination of transverse velocities they suggested using satellite experiments instead (cf. \citealt{refsdal66, gould, gouldpar}).

The model used in this study is the following: at any given time $t$ the observed flux $F(t)$ of a lensed quasar is determined by the quasar intrinsic flux $F_0$ at the time $t-\tau$ and magnification factor $\mu$ corresponding to the observer position $(\tilde{\bf r}, t)$, with gravitational lensing time delay $\tau$ and magnification $\mu$ being some functions of the lensing configuration:
\begin{equation}
	F(t)=F_0(t-\tau(\tilde{\bf r}, t))\times\mu(\tilde{\bf r}, t) \label{Ft}
\end{equation}
The observer moves with the Solar system at an unknown constant projected velocity $\tilde{\bf v}$ and also rotates around the Sun in a well-known annual motion $\tilde{\bf r}_e(t)$ as projected onto the observer's plane defined as the plane which contains the Sun and is orthogonal to the line connecting some fixed points in the source and lens planes:
\begin{equation}
	\tilde{\bf r}(t)=\tilde{\bf v} t +\tilde{\bf r}_e(t). \label{rt}
\end{equation}
In the Galactic case, this equation is often written in terms of the quantity ${\bf u}(t)$ -- the angular distance between the source and the lens ${\Delta\bphi}$ divided by the Einstein-Chwolson angular radius $\theta_E$:
\begin{equation}
{\bf u}(t)\equiv\frac{{\Delta\bphi}}{\theta_E} = \frac{\tilde{\bf r}(t)}{\tilde{r}_E}=\pi_E\frac{\tilde{\bf v}t + \tilde{\bf r}_e(t)}{1\,\rmn{au}}
\label{gart}
\end{equation}
The last two equalities define the Einstein radius projection onto the image plane $\tilde{r}_E$ and microlensing annual parallax $\pi_E=1\,\rmn{au}/\tilde{r}_E$ \citep{gould00}. The advantage of this formulation is that in the case of a Schwarzschild lens the magnification depends solely on the magnitude of ${\bf u}$, determined by the system kinematics, and on $\theta_E$, determined by the system geometry and lens mass; each of these quantities is therefore associated with distinct degeneracies in the determination of microlensing event parameters. This is not the case for the much more complex lenses responsible for quasar microlensing, where $\tilde{r}_E$ itself can only be used in certain order-of-magnitude estimates.

If the intrinsic variability can be neglected, time delays -- which are not independent of  the magnification factors -- drop out of the analysis. This seems to be the case with QSO2237+0305 as the time delay differences between the images expected in this system are small \citep{schneiderondelays}, while the observed light curves show rich structure and do not correlate with each other. Therefore, it is sensible to assume that the variability is dominated by microlensing and put $F_0=\rmn{const}$. In this case we can consider a region centered at some zero point $({\bf r}_0, t_0)$ small enough for the Taylor series expansion 
\begin{equation}
\mu(\tilde{\bf r}, t)=\mu(\tilde{\bf r}_0, t_0)+ (\tilde{\bf r}-\tilde{\bf r}_0){\bf \nabla}\mu + (t-t_0)\frac{\partial\mu}{\partial t}+{\cal R} \label{museries}
\end{equation}
to give a negligible remainder, ${\cal R}$.

As long as intrinsic variations are not involved it is more convenient to deal with astronomical magnitudes. Given the times of observations $t_i$ and corresponding quasar magnitudes $m_i$ we can attempt to estimate the parameters of the representation similar to~(\ref{museries}) by fitting it to the data:
\begin{eqnarray}
m_i-\Delta m_i - {\cal R}_i=m_0 + T(t_i-t_0)+ \hspace{2.7cm} \label{XYT}\\
\hspace{2.5cm}X(\tilde{x}_e(t_i)-\tilde{x}_e(t_0))+Y(\tilde{y}_e(t_i)-\tilde{y}_e(t_0)) \nonumber 
\end{eqnarray}
where $\Delta m_i$ is the actual error of the $i$-th measurement, $X$ and $Y$ are the components of the gradient on the observer's plane while $T$ is the sum of partial derivative with respect to time (caused by microlens proper motions) and `drag term' due to the constant motion of the Solar system:
\begin{equation}
T=\frac{\partial m}{\partial t} + \tilde{\bf v}\cdot{\bf\nabla}m. \label{T}
\end{equation}
All the derivatives correspond to the zero point $({\bf r}_0, t_0)$.

Since the Earth's motion projected onto the observer's plane is an ellipse, the number of parameters fit by~(\ref{XYT}) is four -- two for the straight line corresponding to the average motion of the Solar system and two for the amplitude and phase of the sinusoid due to the Earth's annual motion projected onto the gradient vector of the magnification map. This the minimal number of parameters needed within this model. We will see in the next section that the quality of OGLE data available for QSO2237+0305 is such that a four-parameter fit fully describes it within observational errors. This means that it is not possible to estimate the remainder in a proper way and thereby test the validity of the model.

The major question is whether the partial derivative with respect to $t$ in~(\ref{T}) can be neglected. Although this assumption appears to be somewhat customary throughout many microlensing studies, random motion of stars in the lensing galaxy with respect to each other means that it cannot hold precisely and was in fact shown to be an oversimplification \citep{kundic}. However, the average of this term among different patches of any single light curve should be zero and this can, in accordance with common sense, serve  as a definition of the bulk velocity of microlenses. Therefore, we expect a {\it correlation} between measured values of $T$ and $(X, Y)$ with coefficients being the velocity components:
\begin{equation}
T^{\prime}_j=\tilde{v}_x\cdot X_j + \tilde{v}_y\cdot Y_j \label{condeq}
\end{equation}
Therefore, we need at least three sets of $(T, X, Y)$ to obtain a sensible estimate of $\tilde{\bf v}$.

The velocity thus obtained represents the sum of the transverse velocity in the observer-lens-source system as a whole, $\tilde{\bf v}_0$, and the projected local bulk velocity of the microlenses at the point where the $k$-th macroimage is formed with respect to the lens center of mass $\tilde{\bf u}_k$ transformed by the macrolensing matrix $\hat{A}_k$ acting at this point (see (B9, B10) of \citealt{kayserrefsdalstabell}):
\begin{equation}
	\tilde{\bf v}_k=\tilde{\bf v}_0 + \hat{A}_k\tilde{\bf u}_k
\label{vlocal}
\end{equation}
The $\hat{A}_k$ matrices are obtained by the macrolensing modeling and are usually well-known.

Macrolensing in the observed systems generally tends to magnify images and therefore $\hat{A}$ matrices effectively contract the vectors they act upon, while the magnitude of the velocities of the microlenses is expected to be smaller than the magnitude of the random velocities of the galaxies themselves. Therefore one expects the velocity estimates $\tilde{\bf v}_k$ not to vary much from one image image to another and similarity of these values would provide a strong argument in favour of the validity of the model adopted. The arithmetic mean can then be used as an estimate for $\tilde{\bf v}_0$ which effectively means requiring $\sum\hat{A}_k\tilde{\bf u}_k\approx 0$. Except for possible differences in weighting, this is equivalent to mixing all the images together and fitting one single $\tilde{\bf v}$ in~(\ref{condeq}) for all available $(T, X, Y)$.

Whenever $\hat{A}\tilde{\bf u}$ cannot be considered negligible compared to $\tilde{\bf v}_0$ -- either due to a chance which exists for the overall velocity projection $\tilde{\bf v}_0$ to be low or because of the high values for the individual bulk velocities $\tilde{\bf u}_k$ -- additional information on the latter must be invoked, such as the model of the expected rotation or other galactic-scale pattern of star motion in the lensing galaxy with a subsequent fitting for~(\ref{condeq}) using~(\ref{vlocal}). 

Another possibility which may be present in this case is to utilize the symmetry observed in many gravitationally lensed systems which often suggests that the average of bulk velocities should be close to zero: $\sum\tilde{\bf u}_k\approx 0$. In this case the estimate for the average velocity is
\begin{equation}
	\tilde{\bf v}_0=\left(\sum\limits_k\hat{A}_k^{-1}\right)^{-1}\sum\limits_j\hat{A}_j^{-1}\tilde{\bf v}_j
\label{v0est}
\end{equation}
while individual bulk velocities are clearly
\begin{equation}
\tilde{\bf u}_k=\hat{A}_k^{-1}(\tilde{\bf v}_k-\tilde{\bf v}_0)
\label{uest}
\end{equation}
We now apply the above considerations to the gravitationally lensed quasar QSO2237+0305.

\section{Application to QSO2237+0305}

\begin{figure}
\hspace{0cm}\includegraphics[width=70mm, angle=270]{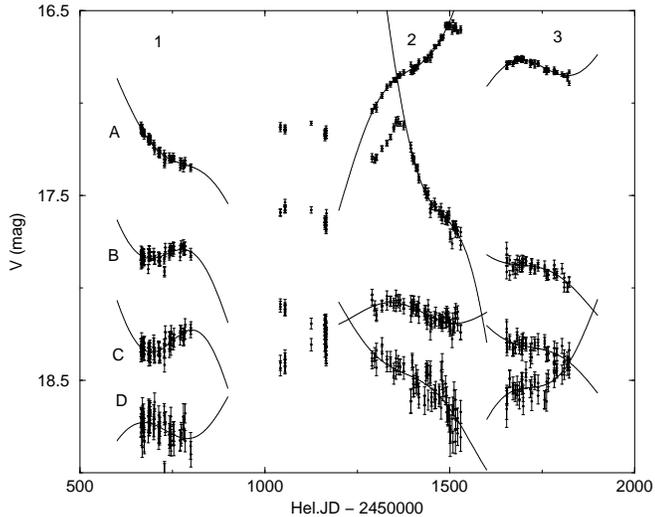}
  \caption{Observed light curve of the four images of the gravitationally lensed quasar QSO2237+0305 obtained by OGLE collaboration during OGLE-II phase of the project. Solid lines are the fits with parameters given in Table 1 (see text for details).}
\label{ogleandfit}
\end{figure}

We use the light curves of four images of QSO2237+0305, published by the OGLE collaboration~\citep{oglehuchra}\footnote{These data are publicly available at {\it http://bulge.princeton.edu/\~{ }ogle/ogle2/huchra.html}}. In the OGLE-II phase of the project over two hundred measurements were obtained for each of the four images of the quasar in more than three years of observations (Fig.~\ref{ogleandfit}). The best coverage was obtained over the following three periods of time (Hel.JD - 2450000): 650-800, 1250-1550 and 1650-1850. In each of these periods the light curve of each of the images visually seems to admit representation~(\ref{XYT}), with the exception of A and C images in the second period where distinct peaks, which could be associated with a close encounter with a critical line or point, made us exclude data with $t>1500$ and $t<1380$, respectively, from the analysis of this period.

We use the SLALIB package~\citep{slalib} to compute the Earth's position as a function of time and then project it on to the plane perpendicular to the line of sight toward the quasar, to get $x_e$ and $y_e$. The $Ox$ axis is chosen along the COBE dipole projection onto the same plane~\citep{cobedipole} and $Oy$ is such that $Ox$, $Oy$ and the direction to the quasar $Oz$ is a right-hand basis (this makes $Oxy$ system apparently left-handed when viewed in the image plane). The model is fit to the data by the least-squares method -- i.e. minimizing the sum $\sum (m_0+T(t_i-t_0)+X(x_i-x_0)+Y(y_i-y_0)-m_i)^2$ for each of the images in each of the periods independently.

Results of the fitting are given in table 1. Here we present the values of $T$, $X$ and $Y$ as well as associated errors. The last two columns show the average fitting quality per datum (which square is defined as the sum of residuals squared divided by the number of data points) and the average observational uncertainty reported by the OGLE group. The fits are shown in figure~\ref{ogleandfit} as solid lines.

\begin{table*}
\begin{centering}
\begin{minipage}{160mm}
\caption[center]{Best fit parameters of the representation~(\ref{XYT}) for images A -- D in each of the three periods (Hel.JD - 2450000):\\ 1) 650 -- 800 (A -- D), 2) 1250 -- 1500 (A), 1250 -- 1550 (B, D), 1380 -- 1550 (C) and 3) 1650 -- 1850 (A -- D).}
\begin{tabular}{ccccrcrcrcrcrc}
\hline
\hspace{0.5cm} Image \hspace{0.5cm} & & Period &\hspace{1cm} & T (mag/yr) & & X (mag/au) & & Y (mag/au) & & $\delta m_{fit}$ & & $\delta m_{obs}$\\
\hline
A & & 1 &	& $  0.95 \pm 0.2 $ & & $ 0.057 \pm 0.04 $ & & $ 0.60 \pm 0.11 $ & & 0.014 & & 0.014 \\
  & & 2 &	& $ -1.41 \pm 0.04$ & & $-0.098 \pm 0.01 $ & & $ 0.50 \pm 0.04 $ & & 0.013 & & 0.010 \\
  & & 3 &	& $ -0.37 \pm 0.07$ & & $-0.073 \pm 0.01 $ & & $ 0.37 \pm 0.09 $ & & 0.013 & & 0.010 \\
B & & 1 &	& $  0.91 \pm 0.4 $ & & $ 0.20  \pm 0.08 $ & & $ 0.61 \pm 0.2  $ & & 0.027 & & 0.022 \\
  & & 2 &	& $ -0.03 \pm 0.07$ & & $-0.063 \pm 0.02 $ & & $-0.01 \pm 0.05 $ & & 0.033 & & 0.032 \\
  & & 3 &	& $  0.51 \pm 0.2 $ & & $ 0.002 \pm 0.03 $ & & $-0.31 \pm 0.3  $ & & 0.036 & & 0.036 \\
C & & 1 &	& $  0.87 \pm 0.4 $ & & $ 0.22  \pm 0.08 $ & & $ 0.97 \pm 0.2  $ & & 0.029 & & 0.035 \\
  & & 2 &	& $  3.14 \pm 0.5 $ & & $ 0.34  \pm 0.09 $ & & $ 1.66 \pm 0.3  $ & & 0.028 & & 0.019 \\
  & & 3 &	& $  0.43 \pm 0.2 $ & & $-0.026 \pm 0.02 $ & & $-0.33 \pm 0.2  $ & & 0.028 & & 0.025 \\
D & & 1 &	& $ -0.43 \pm 0.9 $ & & $-0.14  \pm 0.2  $ & & $-0.41 \pm 0.5  $ & & 0.068 & & 0.054 \\
  & & 2 &	& $  0.80 \pm 0.15$ & & $ 0.035 \pm 0.04 $ & & $-0.29 \pm 0.1  $ & & 0.072 & & 0.050 \\
  & & 3 &	& $ -0.95 \pm 0.3 $ & & $ 0.01  \pm 0.03 $ & & $ 0.72 \pm 0.4  $ & & 0.047 & & 0.047 \\
\hline
\end{tabular}
\end{minipage}
\end{centering}
\end{table*}

One can see that the model provides an adequate fit for the data given the observational uncertainties. Unfortunately, this also means that the accuracy of the data does not allow for additional tests, and we must admit that a simple polynomial in $t$ with the same number of parameters can fit the data with almost the same quality. 

We then apply the method presented in the previous section to estimate the transverse velocity of the system. We first combine twelve sets $(T, X, Y)$ all together regardless of what image they belong to. This is the first option mentioned in the previous section valid when transformed bulk velocities $\hat{A}_k\tilde{\bf u}_k$ are either individually negligible or are zero on average.

\begin{figure}
\hspace{0cm}\includegraphics[width=70mm, angle=270]{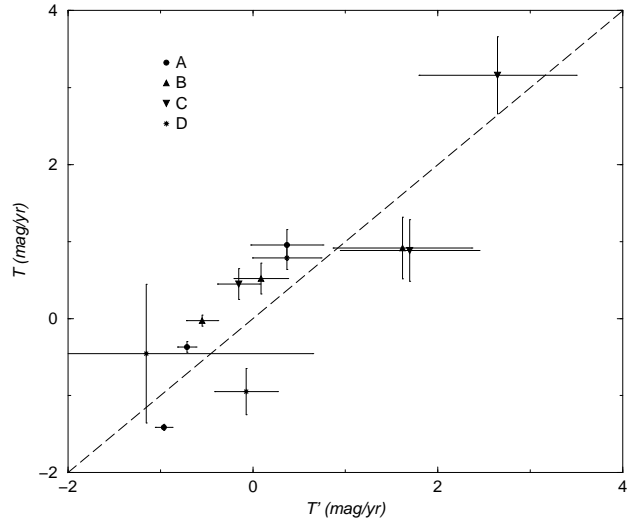}
  \caption{Comparison between calculated $T^{\prime}_j=\tilde{v}_x\cdot X_j + \tilde{v}_y\cdot Y_j$ and actual $T_j$ values of the coefficient $T$ in~(\ref{XYT}). The error bars in ordinate correspond to uncertainties quoted in Table 1, errors in abscissa are the sum of uncertainties in $X$ and $Y$ with coefficients $\tilde{v}_x$, $\tilde{v}_y$.}
\label{TT}
\end{figure}

To make this procedure reflect the fact that different troikas $(T, X, Y)$ have different uncertainties we assigned the weight to each of them inversely proportional to the square of the average fitting quality from Table 1. In fact an unweighted least-squares estimate gives the values of $\tilde{v}_x, \tilde{v}_y$ very close (within formal uncertainties) to those of the weighted one.

The best fit obtained is $\tilde{\bf v}=(41\pm 7, -1\pm 1.5)\,\rmn{km}\,\rmn{s}^{-1}$ ($\tilde{\bf v}=(42\pm 9, -2\pm 2)\,\rmn{km}\,\rmn{s}^{-1}$ in the unweighted case). Fig.~\ref{TT} compares the values of $T_j$ (ordinate) and $T^{\prime}_j=\tilde{v}_x\cdot X_j + \tilde{v}_y\cdot Y_j$ (abscissa). As we have a two-parameter fit here this figure should not be viewed as a standard data spread `cloud'. This is rather the best projection of the data points onto one single plane determined by $\tilde{\bf v}$. The fitted values seem to be stable to subsampling of the $\{(T, X, Y)_j ; j=\overline{1,12}\}$ set. 

The value of the Solar system velocity with respect to the background radiation found by the COBE satellite \citep{cobedipole} projected onto the observer's plane is $\tilde{\bf v}_{COBE}=(56.0\pm 0.5, 0\pm 0.7)\,\rmn{km}\,\rmn{s}^{-1}$. Subtracting this velocity from the value we obtained we get the transverse velocity of the quasar with respect to the lensing galaxy, projected onto the observer plane: $\tilde{\bf v}_{Q-G}^{\perp}=(-15\pm 7, -1\pm 2)\,\rmn{km}\,\rmn{s}^{-1}$ ($\tilde{\bf v}_{Q-G}^{\perp}=(-14\pm 9, -2\pm 3)\, \rmn{km}\,\rmn{s}^{-1}$ in the unweighted case). Taking into account the difference in distances to the lensing galaxy ($z_L\approx 0.0394$) and quasar ($z_S\approx 1.695$) \citep{huchra}, the effective `lever arm' ratio is about 1:10.5 (in $\Omega=0.3$, $\Omega+\Lambda=1$ cosmology), and therefore the measured velocity corresponds to either $\sim 15\,\rmn{km}\,\rmn{s}^{-1}$ in the lens plane or $\sim 10.5\times (1+z_S)\times 15 \, \rmn{km}\,\rmn{s}^{-1}\approx 420\, \rmn{km}\,\rmn{s}^{-1}$ in the quasar plane.

This estimate is curious in both the direction and magnitude. The direction -- which seems to be very close to that of the Earth's projected motion with respect to the cosmic microwave background -- is not, however, a problem since the velocity is expected to be aligned with the C-D image axis based on an independent quiescence argument \citep{wittmao}, and the C-D line is almost exactly aligned with the COBE dipole projection. Yet the magnitude of the transverse velocity seems to be unreasonably small because random individual velocities of galaxies have measured dispersion of a few hundreds kilometers per second (e.g. \citealt{tormen, bahcall96}) consistent with theoretical expectations (e.g., \citealt{moscardini, bahcall94})
thus making a probability to obtain a value of a few dozen kilometers per second by chance very low (for a Gaussian distribution with one-dimensional dispersion of $\sigma_{\tilde{v}}=300\,\rmn{km}\,\rmn{s}^{-1}$ the probability $P$ of the velocity being less than or equal to $\tilde{v}=15 \,\rmn{km}\,\rmn{s}^{-1}$ is $P\approx\tilde{v}^2/(2\sigma_{\tilde{v}}^2)\sim 10^{-3}$). 

This value is also small on the scale of typical velocities of stars within galaxies which means that the assumption of $\hat{A}\tilde{\bf u}$ to be negligible under which it was obtained is unlikely to be true. We therefore attempted to obtain individual velocities for all four macroimages of the quasar. Only three sets $(T, X, Y)_j$ -- the minimal number to make any sensible estimate in this case -- were available for this task and therefore the uncertainties cannot be small. The individual velocities and their formal uncertainties obtained are $\tilde{\bf v}_A=(61\pm 22, 2\pm 4)\,\rmn{km}\,\rmn{s}^{-1}$, $\tilde{\bf v}_B=(32\pm 20, -4\pm 6)\,\rmn{km}\,\rmn{s}^{-1}$, $\tilde{\bf v}_C=(70\pm 110, -7\pm 23)\,\rmn{km},\rmn{s}^{-1}$ and $\tilde{\bf v}_D=(36\pm 4, -6.9\pm 0.5)\,\rmn{km}\,\rmn{s}^{-1}$. 

The configuration of the images seen in QSO2237+0305 shows a high degree of symmetry. We can therefore assume that the sum of individual velocities $\tilde{\bf u}$ should be close to zero. This suggests to use the second method sketched in the previous section. To apply it, we use the macrolensing model of \citet{schmidt} for this lensing system to construct microlensing transformation matrices $\hat{A}_k$ for the images. By requiring $\sum \tilde{\bf u}_k$ to be zero, we get the value of $\tilde{\bf v}_0=(48\pm 10, -2\pm 2)\,\rmn{km}\,\rmn{s}^{-1}$, while for the random velocities we obtain $\tilde{\bf u}_A=(42 \pm 20, -4 \pm 5)\,\rmn{km}\,\rmn{s}^{-1}$, $\tilde{\bf u}_B=(-49 \pm 17, -12 \pm 7)\,\rmn{km}\,\rmn{s}^{-1}$, $\tilde{\bf u}_C=(20 \pm 40, -0.5 \pm 20)\,\rmn{km}\,\rmn{s}^{-1}$, $\tilde{\bf u}_D=(-13 \pm 5, 16 \pm 2)\,\rmn{km}\,\rmn{s}^{-1}$; the uncertainties quoted do not include the uncertainty of the macrolensing model (i.e., those of $\hat{A}_k$).

One can see that both the projected average velocity and random bulk velocities at the image locations are very low. We are therefore inclined to admit that our method described above fails when applied to this system, and the parallax signal which must be present in the light curve is not detected. Deviations of the light curves from straight lines are therefore due to nonlinearity in the magnification map and are not caused by the Earth's annual motion --- that is the $\cal R$ term of equation~(\ref{museries}) is not negligible. In the next section we describe what sort of systems the parallax effect is most likely to be observed in.

\section{Where might the effect be found?}

To understand the conditions that would allow one to detect the annual parallax effect in multiply imaged quasars, let us again rewrite expansion~(\ref{museries}) about some given point (for simplicity, let it denote the coordinate system origin) under the assumption that the derivatives with respect to time are zero and with the remainder in the Lagrange form:
\begin{equation}
	m=m_0+(\tilde{\bf v}t+\tilde{\bf r}_e){\bf\nabla}m + (\tilde{\bf v}t+\tilde{\bf r}_e)^2D_2
\label{museries2}
\end{equation}
Here the first derivative is evaluated at the centre of the expansion while $D_2$ represents a sum of second derivatives at a point between the centre and $\tilde{\bf v}t + \tilde{\bf r}_e$ with certain coefficients of order unity. From this point on the following order-of-magnitude reasoning can be adopted. The natural scale for the $(\tilde{\bf v}t + \tilde{\bf r}_e)$ is the Einstein-Chwolson radius projection onto the observer's plane, $\tilde{r}_E$. With lengths expressed in these units, the gradient (which equals ${\bf\nabla}m\tilde{r}_E$) is typically a quantity of order unity and so is $D_2\tilde{r}^2_E$.

The strength of the parallax signal is determined by ${\bf r}_e{\bf\nabla}m$ with $r_e$ being an oscillating quantity with the amplitude of order one astronomical unit and period of one year. While $\tilde{\bf r}_e$ oscillates, the  $\tilde{\bf v}t$ term does not, and therefore the ratio of these quantities is optimally measured over a time span of about a year; its maximum value is of order of the Earth's orbital velocity over the transverse velocity in the system $\tilde{v}$. The latter is not known for actual gravitationally lensed systems but it is expected to be about a few hundred kilometres per second --- that is, much larger than the Earth's orbital velocity. Therefore we may now write an approximate expression for $m$:
\begin{eqnarray}
\lefteqn{m\approx m_0+\frac{\tilde{\bf v}t + \tilde{\bf r}_e}{\tilde{r}_E} [\tilde{r}_E{\bf\nabla}m] + \left(\frac{\tilde{\bf v}t}{\tilde{r}_E}\right)^2 [\tilde{r}_E^2D_2] \label{muoom}}
\\
\hspace{0cm} & =m_0 + \frac{\tilde{\bf v}t + \tilde{\bf r}_e}{1\,\rmn{au}}\tilde{\pi}_E[\tilde{r}_E{\bf\nabla}m]  + \left(\frac{\tilde{\bf v}t}{1\,\rmn{au}}\right)^2\tilde{\pi}_E^2 [\tilde{r}_E^2D_2]
\nonumber
\end{eqnarray}
As has already been noted, the quantities in the square brackets are supposed to be of order unity, sensible values of $t$ are about half a year and $\tilde{v}$ is about a dozen times the Earth orbital velocity -- $(50-100)\,\rmn{au}/\rmn{yr}$ . The last term in this formula represents contamination and the signal is big relative to it when the following ratio is large:\begin{eqnarray}
\lefteqn{\left(\frac{\tilde{r}_e}{\tilde{r}_E}\left[\tilde{r}_E\nabla m\right]\right)/\left(\left(\frac{\tilde{v}t}{\tilde{r}_E}\right)^2[\tilde{r}_E^2D_2]\right)}\label{sn} \hspace{0cm}\\
 & \sim (1-2)\,\tilde{v}_{300}^{-2}\,\tilde{r}_{E3} \left[\frac{\tilde{r}_E\nabla m}{\tilde{r}_E^2D_2}\right] \sim \tilde{r}_{E3} \sim \tilde{\pi}_{E3}^{-1}
\nonumber
\end{eqnarray}
That is, the effect is likely to be observed in the systems with larger $\tilde{r}_{E3}=\tilde{r}_E/(10^3\,\rmn{au})$ and smaller $\tilde{v}_{300}= \tilde{v}/(300\,\rmn{km}\,\rmn{s}^{-1})$, because the contamination is more rapidly suppresed by increasing $\tilde{r}_E$ (decreasing the microlensing parallax $\tilde{\pi}_E$) compared to the useful signal. This is different from microlensing by Schwarzschild lenses where no expansion is necessary and the second as well as all higher order terms contribute to the useful signal. However, $\tilde{r}_E$ should not be made too large because that would mean a very flat magnification map and consequently a very weak signal, likely to be lost in the observational noise --- the only contaminant which is relevant to the case of a Schwarzschild lens. Flex points (those where the second derivatives vanish) are also highly preferred to points of extrema, though this can hardly be controlled.

The Einstein-Chwolson radius projection $\tilde{r}_E=\tilde{\pi}_E^{-1}\,\rmn{au}$ onto the observer plane here is that of the `natural formalism for microlensing' introduced by \citet{gould00}. In a cosmological context, it is given by
\begin{equation}
	\tilde{r}_E=(1+z_L)\sqrt{\frac{4GM}{c^2}D_{OL}\frac{D_{OS}}{D_{LS}}}
\label{repr}
\end{equation}
which, for a flat Universe amounts to
\begin{eqnarray}
\lefteqn{\tilde{r}_E=\sqrt{\left(1+z_L\right)\frac{4GM}{cH_0}\frac{f(z_L)f(z_S)}{f(z_S)-f(z_L)}}} \hspace{0.2cm} \label{repflat} \\
& = 5.9\times 10^3 \left(\frac{M}{M_{\odot}}\right)^{1/2} \left((1+z_L)\frac{f(z_L)f(z_S)}{f(z_S)-f(z_L)}\right)^{1/2}\,\rmn{au} \nonumber
\end{eqnarray}
(for $H_0=70\,\rmn{km}\,\rmn{s}^{-1}\,\rmn{Mpc}^{-1}$) with (e.g., \citealt{weinberg})
\begin{equation}
f(z)\equiv\int\limits_0^z\frac{\rmn{d}t}{\sqrt{\Omega_0(1+t)^3+(1-\Omega_0)}}
\label{fotz}
\end{equation}
-- a monotonic increasing function of $z$.

The last ratio in~(\ref{repflat}) may be the reason why the effect was not detected in QSO2237+0305 -- close proximity of the lens makes the value of $\tilde{r}_E$ relatively low. Placing the lens closer to the source, increases the value of $\tilde{r}_E$ along with the chances to detect the parallax effect but it cannot be placed too close as this would reduce the magnification and ability to produce multiple images.

Not much can be said about another player in~(\ref{sn}) -- the transverse velocity $\tilde{v}$. It is generally completely unknown and it is for the determination of this quantity that the method described here is intended. One of our suggestions is to keep away from galaxy clusters, where virialized galaxies are expected to have velocities relatively high compared with those of field galaxies. In addition, it should be mentioned that the Solar transverse velocity with respect to the CMBR is of the same order as or even greater then the galaxies' velocity dispersion \citep{bernardeaureview}, and therefore the expected value of the effective transverse velocity is lower in the regions closer to the direction of COBE dipole.

With regards to the quantity in the square brackets in~(\ref{sn}) nothing definite is usually known before the observations are done. However, there is a way to increase this fraction by decreasing the value in denominator. Ceteris paribus, convolution of the magnification map corresponding to a point-like source with an extended profile affects higher order derivatives of the resulting map in a greater degree than lower order ones. Therefore, there is a possibility to suppress the contamination from curvature in the intrinsic magnification map relative to the linear signal by observing the system in a spectral region where the source looks larger -- e.g., in broad lines -- provided that the smoothing scale is of about the same order as the scale of the contaminating signal or larger. This effect will mostly be limited by the scale at which it will start to suppress the linear term greater than allowed by the observational noise.

Now we can formulate a short list of properties that a lensing system should possess to make the detection of the parallax effect and determination of the transverse velocity possible:
\begin{enumerate}
\item the lens should be located relatively far away and therefore so should be the source
\item the lens and the source should be relatively close to each other -- though to an extent that still allows multiple imaging
\item the lens should not be a virialized member of a massive galaxy cluster
\item the system should be located not far from the direction of Solar motion relative to CMBR
\item either the configuration of the system should be highly symmetric or there must be some additional information available that could constrain the bulk velocities at image locations \item high degree of symmetry is also an advantage when there is a need to constrain the intrinsic variability of the source as the time delay differences between images are generally low in such systems
\item the lens should be preferably a spiral galaxy with a low value of local velocity dispersion at image locations because otherwise it would be difficult to link the measured velocity of the magnification map with physical velocities of microlenses, which, after all, are the actual values of interest
\item where possible, observations in the continuum should be supplemented by the observations in broad lines or other bands where the source appears larger.
\end{enumerate}

In addition, standard observational considerations apply -- the images should be reasonably bright to allow accurate photometry while there also should be a way to gather some additional information on the lensing galaxy. At the moment data of reasonable accuracy and extent are only available for QSO2237+0305 \citep{oglehuchra} and an attempt to determine the transverse velocity in this system presented here was not successful. However, as more gravitationally lensed systems are monitored for microlensing, some of them may be expected to fulfil the conditions listed above, making our method applicable. We suggest that observers might usefully focus on such systems.

\section*{Acknowledgments}
AVT is supported by IPRS and IPA from the University of Sydney. We thank the referee for stimulating comments and questions.

\label{lastpage}

\end{document}